\documentclass[5p]{elsarticle}

\usepackage{lineno}
\usepackage{xcolor}
\usepackage{graphicx} 
\usepackage{subfigure}
\usepackage{dcolumn} 
\usepackage{bm}
\usepackage{float}
\usepackage{hyperref}
\usepackage{amsmath}

\begin{document}
\title{Impact of globally spin-aligned vector mesons on the search for the chiral magnetic effect in heavy-ion collisions}
\author{Diyu Shen, Jinhui Chen}
\address{Key Laboratory of Nuclear Physics and Ion-beam Application (MOE), Institute of Modern Physics, Fudan University, Shanghai 200433, China}
\address{Shanghai Research Center for Theoretical Nuclear Physics, NSFC and Fudan University, Shanghai 200438, China}
\author{Aihong Tang \corref{cor1}}
\address{Brookhaven National Laboratory, Upton, New York 11973}
\ead{aihong@bnl.gov}
\cortext[cor1]{Corresponding author}
\author{Gang Wang}
\address{Department of Physics and Astronomy,
  University of California, Los Angeles, California 90095, USA}

\begin{abstract}
In high-energy heavy-ion collisions, the chiral magnetic effect (CME) is predicted to arise from the interplay between the chirality imbalance of quarks in the nuclear medium and the intense magnetic field, and will cause a charge separation along the magnetic field direction. 
While the
search for the CME is still ongoing in experiments at  Relativistic Heavy Ion Collider (RHIC) and the Large Hadron Collider (LHC),  the CME-sensitive observables need to be scrutinized to exclude the non-CME contributions.
In this work, we examine the influence of globally spin-aligned $\rho$ mesons on the   $\gamma_{112}$ correlator, the $R_{\Psi_2}(\Delta S)$ correlator, and the signed balance functions, via a toy model and a multiphase transport model (AMPT).
The global spin alignment of vector mesons could originate from non-CME mechanisms in heavy-ion collisions, and is  characterized by the 00-component of the spin density matrix, $\rho_{00}$. We find that the CME observables show similar dependence on $\rho_{00}$, and could receive  a positive (negative) contribution from $\rho$-decay pions,  if the $\rho_{00}$ of $\rho$ mesons is larger (smaller) than 1/3. Since pions are the most abundant particles in such collisions, the $\rho_{00}$ measurements for $\rho$ mesons become crucial to the interpretation of the CME data. 
\end{abstract}

\maketitle

\section{Introduction}
Heavy-ion collision experiments provide a unique test ground for  quantum chromodynamics (QCD), as the creation of the quark gluon plasma (QGP) brings about many novel phenomena in the strong interaction. In this paper, we focus on two  such phenomena, the chiral magnetic effect (CME) and the global spin alignment of vector mesons, and study how the latter could affect the experimental search for the former. 
The CME postulates that in non-central collisions,  chiral quarks in the QGP interact with the intense magnetic field, and form an electric current along the magnetic field direction~\cite{Kharzeev:2004ey}, where the chirality imbalance could emerge from the chiral anomaly in QCD~\cite{Kharzeev:1998kz,Kharzeev:1999cz,Kharzeev:2020jxw}, and the magnetic field is generated by the energetic  nuclear fragments~\cite{Voronyuk:2011jd,Deng:2012pc}. The confirmation of the CME would reveal the topological structure of QCD, and indicate ${\cal P}$ and ${\cal CP}$ violation in the strong interaction~\cite{Kharzeev:1998kz,Kharzeev:1999cz,Kharzeev:2020jxw}. 
Experiments at RHIC and the LHC have been searching for the CME over the past two decades without a definite conclusion~\cite{STAR:2009wot,STAR:2013ksd,STAR:2014uiw,STAR:2021mii,STAR:2022ahj,ALICE:2012nhw,CMS:2016wfo,CMS:2017lrw,ALICE:2017sss}, because of the incomplete understanding of the underlying backgrounds~\cite{Liao:2010nv,Pratt:2010zn,Wang:2016iov,Feng:2018chm,Tang:2019pbl,Wu:2022fwz}. 

The major CME-sensitive observables include the $\gamma_{112}$ correlator~\cite{Voloshin:2004vk}, the $R_{\Psi_2}(\Delta S)$ correlator~\cite{Magdy:2017yje}, and the signed balance functions~\cite{Tang:2019pbl}.
A recent study~\cite{Choudhury:2021jwd} has verified the equivalence in the core
components among these methods, and ascertains that they have similar sensitivities to the true CME signal, as well as the background due to the collective motion (elliptic flow) of the collision system.
Besides the flow-related background, Ref.~\cite{Tang:2019pbl} points out that the signed balance functions could acquire non-trivial contributions from globally spin-aligned vector mesons, which heralds the impact of the global spin alignment of resonances on all these observables.

The global spin alignment of vector mesons originates from the polarization of constituent quarks, which can be caused by the spin-orbital coupling~\cite{Liang:2004ph,Liang:2004xn,Becattini:2013vja, Yang:2017sdk}, strong electromagnetic fields~\cite{Sheng:2019kmk,Yang:2017sdk}, the local spin alignment~\cite{Xia:2020tyd,Gao:2021rom}, locally fluctuating axial charge currents~\cite{Muller:2021hpe}, and strong vector meson fields~\cite{Sheng:2019kmk, PhysRevD.105.099903, Sheng:2020ghv, Sheng:2022wsy, Sheng:2022ffb}. Polarized quarks can be manifested in both the global polarization of hyperons and the global spin alignment of vector mesons. Non-trivial hyperon global polarization values have been reported at RHIC and the LHC for $\Lambda$($\bar{\Lambda}$)~\cite{STAR:2017ckg,STAR:2018gyt,ALICE:2019onw,STAR:2021beb}, $\Xi$~\cite{STAR:2020xbm}, and $\Omega$~\cite{STAR:2020xbm}. The global spin alignment has also been measured for $K^{*0}$, $\phi$~\cite{STAR:2008lcm,ALICE:2019aid,STAR:2022fan} and $J/\psi$~\cite{ALICE:2022sli}. In particular, unexpectedly large $\rho_{00}$ values for $\phi$ mesons have been disclosed by the STAR Collaboration~\cite{STAR:2022fan}.

In the following sections, we will demonstrate that globally spin-aligned vector mesons could provide a finite contribution to each of the CME-sensitive observables under study. The effect can be qualitatively understood with the aid of analytical derivations, and is further investigated with a toy model and a multiphase transport model. This work suggests that the global spin alignment of $\rho$ mesons is a crucial component in the background estimation for the CME  measurements involving pions.  

\section{The \texorpdfstring{$\gamma_{112}$}{Lg} correlator}

The CME-induced charge separation as well as the other modes of collective motion is  usually studied with the azimuthal angle distribution of final-state particles in the momentum space~\cite{Voloshin:2004vk},
\begin{flalign}
&E\frac{d^3N}{d^3p} = \frac{1}{2\pi}\frac{d^2N}{p_Tdp_Td\mathsf{y}}\left( 1+ 2a_1\sin \Delta \phi+ \sum_{n=1}^{\infty} 2v_n\cos n\Delta \phi  \right),&
\label{Eq: Fourier expansion}
\end{flalign}
where $p_T=\sqrt{p_x^2+p_y^2}$ is transverse momentum, $\mathsf{y}$ is rapidity, and $\Delta \phi$ is the azimuthal angle of a particle relative to the reaction plane (spanned by the impact parameter and the beam momenta). $a_1$ characterizes the charge separation perpendicular to the reaction plane, and the $a_1$ values for positive and negative charges bear opposite signs, i.e., $a_{1,+} = -a_{1,-}$ in a charge-symmetric system. $v_n$ denotes the $n^{\rm th}$-harmonic flow coefficient of final-state particles, and conventionally $v_2$ is called elliptic flow. 

The CME-sensitive observable $\gamma_{112}$~\cite{Voloshin:2004vk} is defined as 
\begin{flalign}
&\gamma_{112} \equiv \left \langle \cos(\phi_\alpha + \phi_\beta - 2\Psi_{\rm RP}) \right \rangle ,&
\end{flalign}
where $\phi_\alpha$ and $\phi_\beta$ are the azimuthal angles of particles $\alpha$ and $\beta$, respectively, and $\Psi_{\rm RP}$ represents the reaction plane. The bracket means averaging over all particles and all events.  The difference in $\gamma_{112}$ between opposite-sign (OS) and same-sign (SS) pairs is supposed to reflect the CME signal, i.e.,
\begin{flalign}
&\Delta \gamma_{112} \equiv \gamma_{112}^{\rm{OS}} - \gamma_{112}^{\rm{SS}} \approx 2a_1^2.&
\end{flalign}
However, $\Delta \gamma_{112}$ is contaminated with backgrounds, e.g., decay daughters of flowing resonances~\cite{Voloshin:2004vk}. As a result, the expansion of $\gamma_{112}^{\rm{OS}}$ contains finite covariance terms. For example, the $\rho$ mesons decay to charged pions:
\begin{flalign}
&\gamma_{112}^{\rm{OS}} = \left \langle \cos(\phi_+ + \phi_- - 2\Psi_{\rm RP}) \right \rangle \notag \\ 
&= \left \langle \cos \Delta \phi_+ \right \rangle \left \langle \cos \Delta \phi_- \right \rangle + \frac{N_\rho}{N_+N_-}\rm{Cov}( \cos \Delta \phi_+, \cos \Delta \phi_- ) \notag \\
& - \left \langle \sin \Delta \phi_+ \right \rangle \left \langle \sin \Delta \phi_-  \right \rangle - \frac{N_\rho}{N_+N_-}\rm{Cov}( \sin \Delta \phi_+, \sin \Delta \phi_- ),&
\label{Eq: gammaOS}
\end{flalign}
where Cov$(a,b)$ denotes the covariance of variables $a$ and $b$. $N_\rho$ is the yield of $\rho$ mesons, and $N_+$ and $N_-$ are the numbers of $\pi^+$ and $\pi^-$, respectively. Note that the covariance terms in $\gamma_{112}^{\rm{SS}}$ could also be finite, owing to mechanisms such as transverse momentum conservation~\cite{Pratt:2010zn}, but this effect should be cancelled in $\Delta \gamma_{112}$. 

\begin{figure}[htbp]
\centering
\vspace*{-0.1in}
\includegraphics[width=0.30\textwidth]{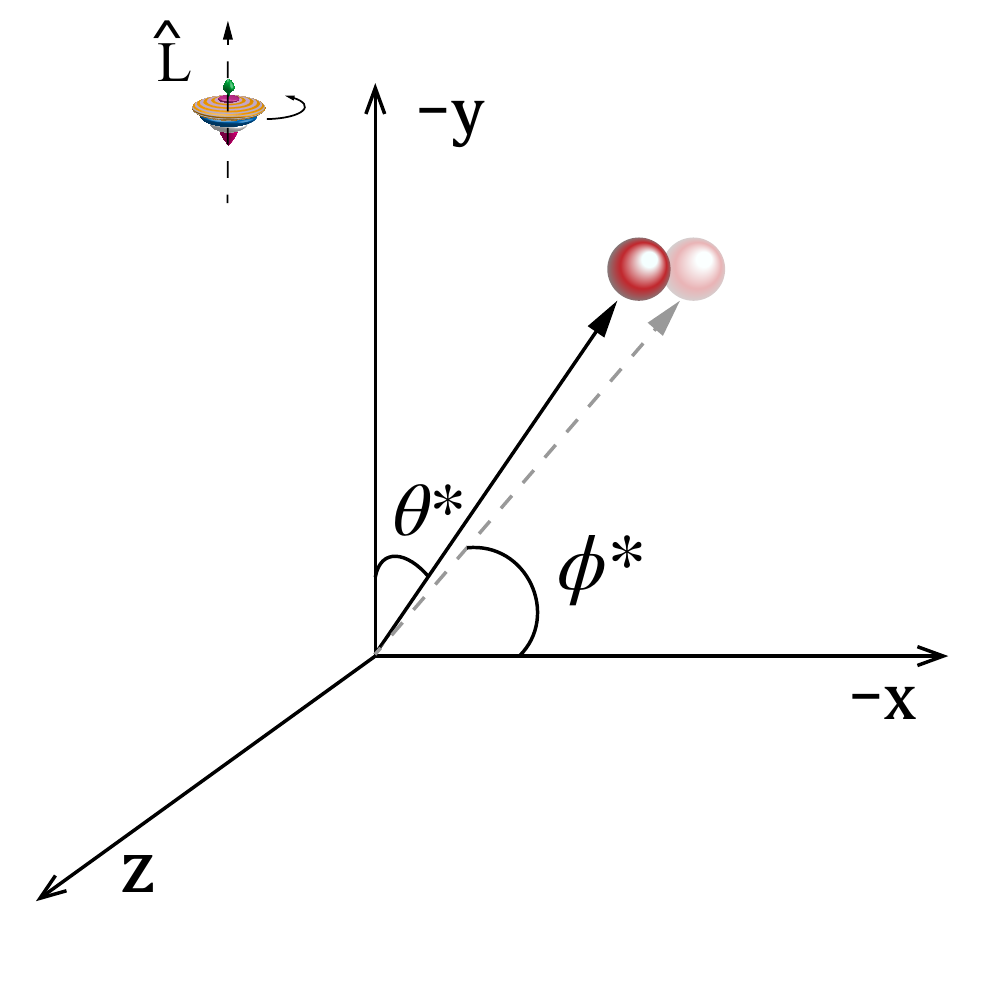}
\caption{\label{fig:Coordinate}
Illustration of pion emission (black solid arrow) in the rest frame of the parent $\rho$ meson. The dashed arrow shows the projection in the $x$-$y$ plane.}
\end{figure}

In addition to elliptic flow, another property of vector mesons, the global spin alignment, could also contribute a background to $\Delta \gamma_{112}$. In the decay of $\rho \rightarrow \pi^{+} + \pi^{-}$, the emission angle of $\pi^\pm$ can be expressed as~\cite{Schilling:1969um} 
\begin{flalign} 
&\frac{dN}{d\cos\theta^*} = \frac{3}{4} \left[ (1-\rho_{00}) + (3\rho_{00}-1)\cos^2\theta^* \right], &
\label{Eq:Spin-theta}
\end{flalign}
where $\theta^*$ is the angle between the pion momentum in the $\rho$ rest frame and the angular momentum $\hat{L}$ of the collision system. In this work, we assume the directions of $\hat{L}$ and the magnetic field are both perpendicular to the reaction plane. $\rho_{00}$ is the 00-component of the spin density matrix, and quantifies the global spin alignment. Projected to the transverse ($x$-$y$) plane, Eq.~(\ref{Eq:Spin-theta}) becomes
\begin{flalign} 
&\frac{dN}{d\phi^*} = \frac{1}{2\pi}\left[1-\frac{1}{2}(3\rho_{00}-1)\cos 2\phi^* \right], &
\label{Eq:Spin-phi}
\end{flalign}
where $\phi^*$ is the azimuthal angle of the decay product in the $\rho$ rest frame as sketched in Fig.~\ref{fig:Coordinate}. Then, the covariance terms in Eq.~(\ref{Eq: gammaOS})  can be calculated in the $\rho$ rest frame as 
\begin{flalign}
&\rm{Cov}(\cos \phi_+^*, \cos \phi_-^* )
  = -\left \langle \cos^2 \phi_+^* \right \rangle + \left \langle \cos \phi_+^* \right \rangle^2  \notag \\
 &= - \frac{1}{2} + \frac{1}{8}(3\rho_{00}-1) ,  \label{Eq:Covcc} \\
&\rm{Cov}( \sin \phi_+^*, \sin\phi_-^* ) 
  = -\left \langle \sin^2 \phi_+^* \right \rangle + \left \langle \sin \phi_+^* \right \rangle^2  \notag \\
 &= - \frac{1}{2} - \frac{1}{8}(3\rho_{00}-1).&  \label{Eq:Covss}
\end{flalign} 
Here we use $\phi_-^* = \phi_+^* +  \pi$. In the absence of the CME, charged pions have $\left \langle \cos \phi_+^* \right \rangle = \left \langle \cos \phi_-^* \right \rangle$ and $\left \langle \sin \phi_+^* \right \rangle = \left \langle \sin \phi_-^* \right \rangle$,   and thus  the $\Delta \gamma_{112}$ in the $\rho$ rest frame can be written as 
\begin{flalign}
&\Delta \gamma_{112}^* = \frac{N_\rho}{N_+N_-}\frac{3\rho_{00}-1}{4},&
\label{Eq:9}
\end{flalign}
which represents a finite background when $\rho_{00}$ deviates from $1/3$.

Essentially, Eq.~(\ref{Eq:Spin-phi}) manifests the elliptic flow of decay products in the $\rho$ rest frame, $v_2^* = -(3\rho_{00}-1)/4$, and therefore Eq.~(\ref{Eq:9}) can be viewed as $\Delta \gamma_{112}^{*} \propto -v_2^*$. In the laboratory frame, Eqs.~(\ref{Eq:Covcc}) and (\ref{Eq:Covss}) should be scaled by  factors of $f_c$ and $f_s$, respectively, due to the Lorentz boost of the $\rho$ meson. In general,  $f_c$ and $f_s$ are different because of the anisotropic motion ($v_2^\rho$) of $\rho$ mesons. One can thus expand $f_c$ and $f_s$ with $v_2^\rho$, 
\begin{flalign}
&f_c = f_0 + \sum_{n=1}^{\infty} c_n (v_2^\rho)^n, \\
&f_s = f_0 + \sum_{n=1}^{\infty} s_n (v_2^\rho)^n, &
\end{flalign}  
where $f_0$, $c_n$ and $s_n$ depend on the spectra of $\rho$ mesons.
Hence, the contribution of decay pions to $\Delta \gamma_{112}$ in the laboratory frame can be expressed as
\begin{flalign}
&\Delta \gamma_{112} = \frac{N_\rho}{N_+N_-}\left [ \frac{1}{8}(f_c+f_s)(3\rho_{00}-1)-\frac{1}{2}(f_c -f_s) \right ] \notag \\
&= \frac{N_\rho}{8N_+N_-}\left[ 2f_0 + \sum_{n=1}^{\infty} (c_n+s_n) (v_2^\rho)^n \right](3\rho_{00}-1)\notag \\
& -\frac{N_\rho}{2N_+N_-}\sum_{n=1}^{\infty} (c_n-s_n) (v_2^\rho)^n .&
\label{Eq:Gamma_lab}
\end{flalign} 
At a given $v_2^\rho$, we expect the $\Delta \gamma_{112}$ measurement involving $\rho$-decay pions to have a linear dependence on the $\rho_{00}$ of $\rho$ mesons. 

\begin{figure}[htbp]
	\centering
	\vspace*{-0.1in}
{\includegraphics[scale=0.35]{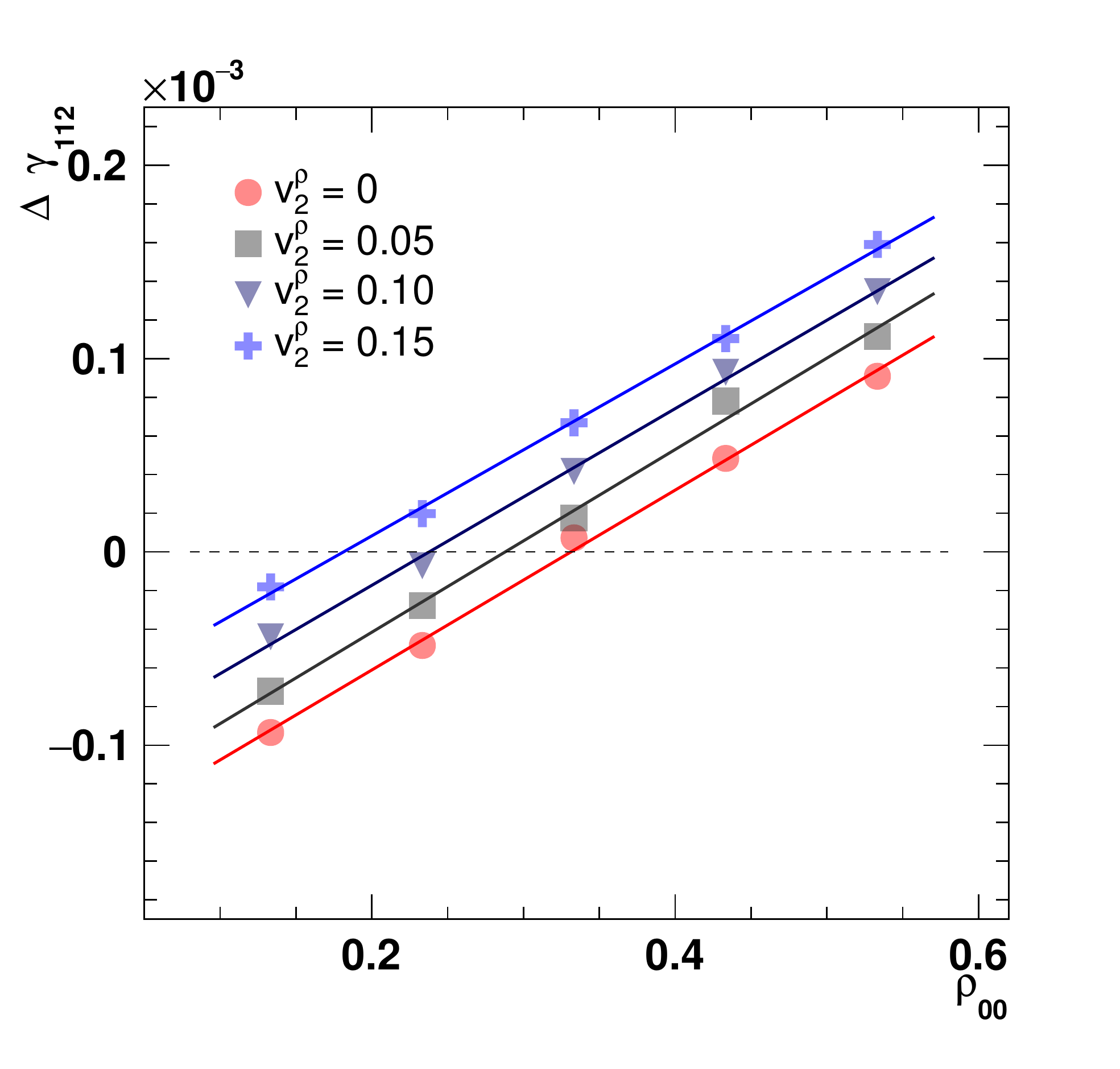}} 	
\caption{Toy model simulations of the $\pi$-$\pi$ $\Delta \gamma_{112}$ correlation as a function of $\rho$-meson $\rho_{00}$ with various inputs of $v_2^\rho$.  Linear fits are applied to guide eyes.}
	\label{fig:Gamma_toy}
\end{figure}
We first test the aforementioned idea with toy model simulations without the CME. Each event contains 195 $\pi^+$ and 195 $\pi^-$, with 33 pairs of them from $\rho$ decays. For simplicity, the $v_2$ and $v_3$ values of primordial pions are set to zero. The spectrum of primordial pions obeys the Bose-Einstein distribution,
\begin{flalign}
&\frac{dN_{\pi^\pm}}{dm_T^2} \propto \frac{1} {e^{m_T/T_{\rm{BE}}} - 1},&
\end{flalign} 
where $T_{\rm{BE}}=212$ MeV is set to match the experimentally observed $\left \langle p_T \right \rangle=$ 400 MeV~\cite{STAR:2008med}. The spectrum of $\rho$ mesons follows
\begin{flalign}
&\frac{dN_\rho}{dm_T^2} \propto \frac{e^{-(m_T-m_\rho)/T}}{T(m_\rho+T)},&
\end{flalign}
where $T = 317$ MeV is set to match its $\left \langle p_T \right \rangle$ of 830 MeV as observed in data~\cite{STAR:2003vqj}. Pseudorapidity (Rapidity) is
uniformly distributed in the range of $[-1, 1]$ for primordial pions ($\rho$ mesons). $\rho$-meson decays are implemented with PYTHIA6~\cite{Sjostrand:2006za}, and the spin alignment effect is simulated by sampling the decay products according to Eq.~(\ref{Eq:Spin-theta}). 

Figure~\ref{fig:Gamma_toy} shows the simulation results of the $\pi$-$\pi$ $\Delta \gamma_{112}$ correlation from the toy model. Each marker  denotes a different input of $v_2^\rho$. It is confirmed that at a given $v_2^\rho$, $\Delta \gamma_{112}$ increases linearly with $\rho_{00}$. On the other hand, $\Delta \gamma_{112}$ also increases with $v_2^\rho$ at a fixed $\rho_{00}$, exhibiting the convolution of $v_2^\rho$ and $\rho_{00}$ in the background contribution from $\rho$ mesons to $\Delta \gamma_{112}$. In the case of $v_2^\rho =0$ and $\rho_{00}=1/3$, $\Delta \gamma_{112}$ is zero, as expected by Eq.~(\ref{Eq:Gamma_lab}). Note that the global spin alignment effect could give a negative contribution to the $\Delta \gamma_{112}$ measurement if $\rho_{00}$ is smaller than $1/3$. 

\begin{figure}[htbp]
	\centering
	\vspace*{-0.1in}
	{\includegraphics[scale=0.35]{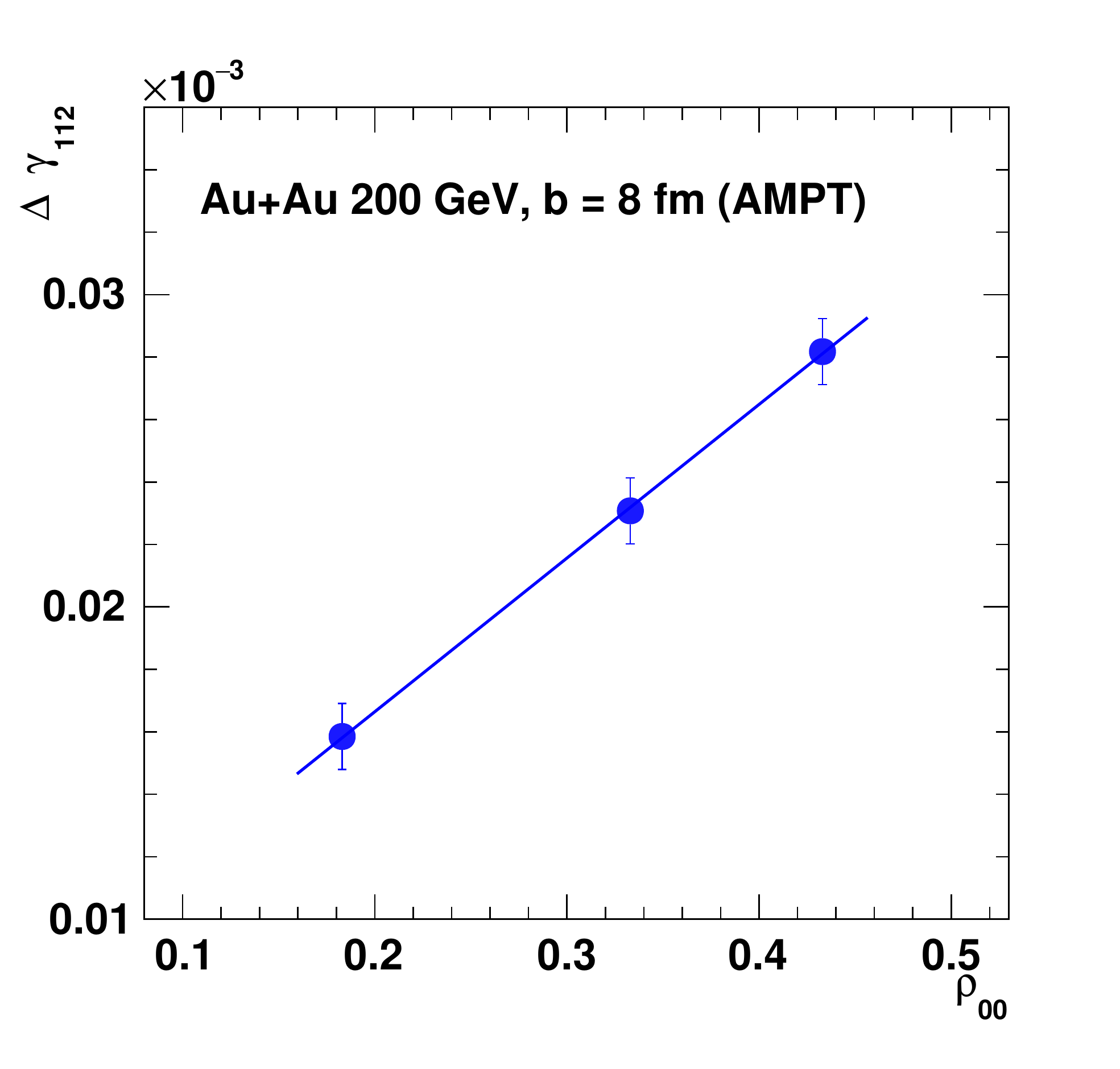}} 	
\caption{AMPT calculations of the $\pi$-$\pi$ $\Delta \gamma_{112}$ correlation as a function of $\rho$-meson $\rho_{00}$ in  Au+Au collisions at $\sqrt{s_{NN}}=200$ GeV with the impact parameter of 8 fm. Solid line represents a linear fit of calculations to guide eyes.}
	\label{fig:Gamma_AMPT}
\end{figure}

We also study this effect with a more realistic model, a multiphase transport (AMPT) model, without the CME and with the spin alignment  implemented by redistributing the momenta of decay products according to Eq.~(\ref{Eq:Spin-theta}). The details of this model can be found in Ref.~\cite{Lan:2017nye,Shen:2021pds}. The selected decay channel is $\rho \rightarrow \pi^+ + \pi^-$. As a qualitative investigation, we only simulate Au+Au collisions at $\sqrt{s_{NN}}=200$ GeV with the impact parameter of 8 fm, and pions are analyzed without any kinematic cut to increase statistics. The $\rho_{00}$ values are set to be $0.18$, $1/3$ and $0.43$, respectively, with a million events for each case. Figure~\ref{fig:Gamma_AMPT} shows the AMPT calculations of $\pi$-$\pi$ $\Delta \gamma_{112}$ as a function of $\rho$-meson $\rho_{00}$. $\Delta \gamma_{112}$ increases linearly with $\rho_{00}$, similar to the  toy model simulations. At $\rho_{00}=1/3$, the positive $\Delta \gamma_{112}$, which is a non-CME background, may come from the positive $v_2^\rho$ and transverse momentum conservation. The slope, $d \Delta \gamma_{112}/d\rho_{00}$, could be different between the toy model and the AMPT model, because of the different $\rho$-meson spectra. 

\section{The \texorpdfstring{$R_{\Psi_2}(\Delta S)$}{Lg} correlator}
Another CME-sensitive observable, the $R_{\Psi_2}(\Delta S)$ correlator~\cite{Magdy:2017yje}, is defined as a double ratio of four distributions,
\begin{flalign}
&R_{\Psi_2}(\Delta S) \equiv \frac{N(\Delta S_{{\rm real}})}{N(\Delta S_{{\rm shuffled}})} / \frac{N(\Delta S^\perp_{{\rm real}})}{N(\Delta S^\perp_{{\rm shuffled}})} ,& \label{Eq:R2} 
\end{flalign}
where
\begin{flalign}
&\Delta S = \left \langle \sin \Delta \phi_+\right \rangle - \left \langle \sin \Delta \phi_-\right \rangle,  
\label{Eq:R3} \\
&\Delta S^\perp = \left \langle \cos \Delta \phi_+\right \rangle - \left \langle \cos \Delta \phi_-\right \rangle, & 
\label{Eq:R4}
\end{flalign}
and $\Delta \phi = \phi - \Psi_2$. $\Psi_2$ denotes the $2^{\rm nd}$-order event plane, an approximation of the reaction plane using the $v_2$ information of produced particles. The bracket means averaging over all particles of interest in an event. The subscripts ``real'' and ``shuffled'' represent real events and charge shuffled events, respectively. Ideally, the CME should cause a concave shape in $R_{\Psi_2}(\Delta S)$, which can be quantified by the width of a Gaussian fit, $\sigma_{R}$. Analytically, $\sigma_{R}$ is related to the widths of the four initial distributions, 
\begin{flalign}
&\frac{S_{\rm concavity}}{\sigma_{R}^2}  = \frac{1}{ \sigma^2(\Delta S_{\rm real})} - \frac{1}{\sigma^2(\Delta S_{\rm shuffled})} - \frac{1}{\sigma^2(\Delta S_{\rm real}^{\perp})} \notag \\
&+ \frac{1}{\sigma^2(\Delta S_{\rm shuffled}^{\perp})}. &
\label{Eq:sigma_R2} 
\end{flalign}
$S_{\rm concavity}$ is 1 (-1) when the $R_{\Psi_2}(\Delta S)$ distribution is convex (concave).

Similar to the case of $\Delta \gamma_{112}$, we first evaluate each term in Eq.~(\ref{Eq:sigma_R2}) in the $\rho$ rest frame, and then use parameters $f_c$ and $f_s$ to describe the Lorentz effect along and perpendicular to the reaction plane, respectively, in the laboratory frame. For example,
$\sigma^2(\Delta S)$ can be expressed as 
\begin{flalign}
&\sigma^2(\Delta S) = \sigma^2(a_1^+) + \sigma^2(a_1^-) - 2 {\rm Cov} (a_1^+, a_1^-) \notag \\
&= \frac{\sigma^2(\sin \Delta \phi_+)}{N_+}  + \frac{\sigma^2 (\sin \Delta \phi_- )}{N_-} \notag \\
&  - \frac{2N_\rho {\rm Cov}( \sin\Delta \phi_+, \sin\Delta \phi_-)}{N_+N_-},& 
\end{flalign}
where $a_1^+$ and $a_1^-$ denote $\langle \sin\Delta \phi_+\rangle$ and $\langle \sin\Delta \phi_-\rangle$, respectively, in an event. After applying Eq.~(\ref{Eq:Spin-phi}), we convert each term into the laboratory frame, 
\begin{flalign}
&\sigma^2(\Delta S_{\rm real}) = f_s\left[ \sigma_s^2 + \frac{N_\rho}{N_+N_-}(1 + \frac{3\rho_{00}-1}{4}) \right], \label{Eq:VarYReal} \\
&\sigma^2(\Delta S_{\rm shuffled}) =  f_s \sigma_s^2,  \label{Eq:VarYShuf}\\
&\sigma^2(\Delta S_{\rm real}^{\perp}) = f_c\left [ \sigma_c^2 +\frac{N_\rho}{N_+N_-}(1 - \frac{3\rho_{00}-1}{4}) \right ], \label{Eq:VarXReal} \\
&\sigma^2(\Delta S_{\rm shuffled}^{\perp}) = f_c \sigma_c^2,& \label{Eq:VarXShuf} 
\end{flalign}
where 
\begin{flalign}
&\sigma_s^2 = \frac{\sigma^2 (\sin \phi_+^* )}{N_+}  + \frac{\sigma^2 (\sin \phi_-^*)}{N_-}, \\
&\sigma_c^2 = \frac{\sigma^2(\cos \phi_+^*) }{N_+}  + \frac{\sigma^2 (\cos \phi_-^*)  }{N_-}. &
\end{flalign}

The sign of $S_{\rm concavity}$ in Eq.~(\ref{Eq:sigma_R2}) is not instantly clear, because $\sigma_c$ and $\sigma_s$, as well as $f_c$ and $f_s$, could be different due to elliptic flow of primordial pions and $\rho$ mesons. Assuming $v_2^\pi$ and $v_2^\rho$ are both zero, we have
\begin{flalign}
&{\rm Sign}(S_{\rm concavity}) = {\rm Sign}\left[ - \frac{N_\rho}{2N_+N-}(3\rho_{00}-1) \right].&
\label{Eq:26}
\end{flalign} 
In this case, if $\rho_{00}$ is smaller (larger) than $1/3$, $S_{\rm concavity}$ becomes 1 (-1), and the $R_{\Psi_2}(\Delta S)$ distribution becomes convex (concave). 

\begin{figure*}[htbp]
	\centering
{\includegraphics[width=0.39\linewidth]{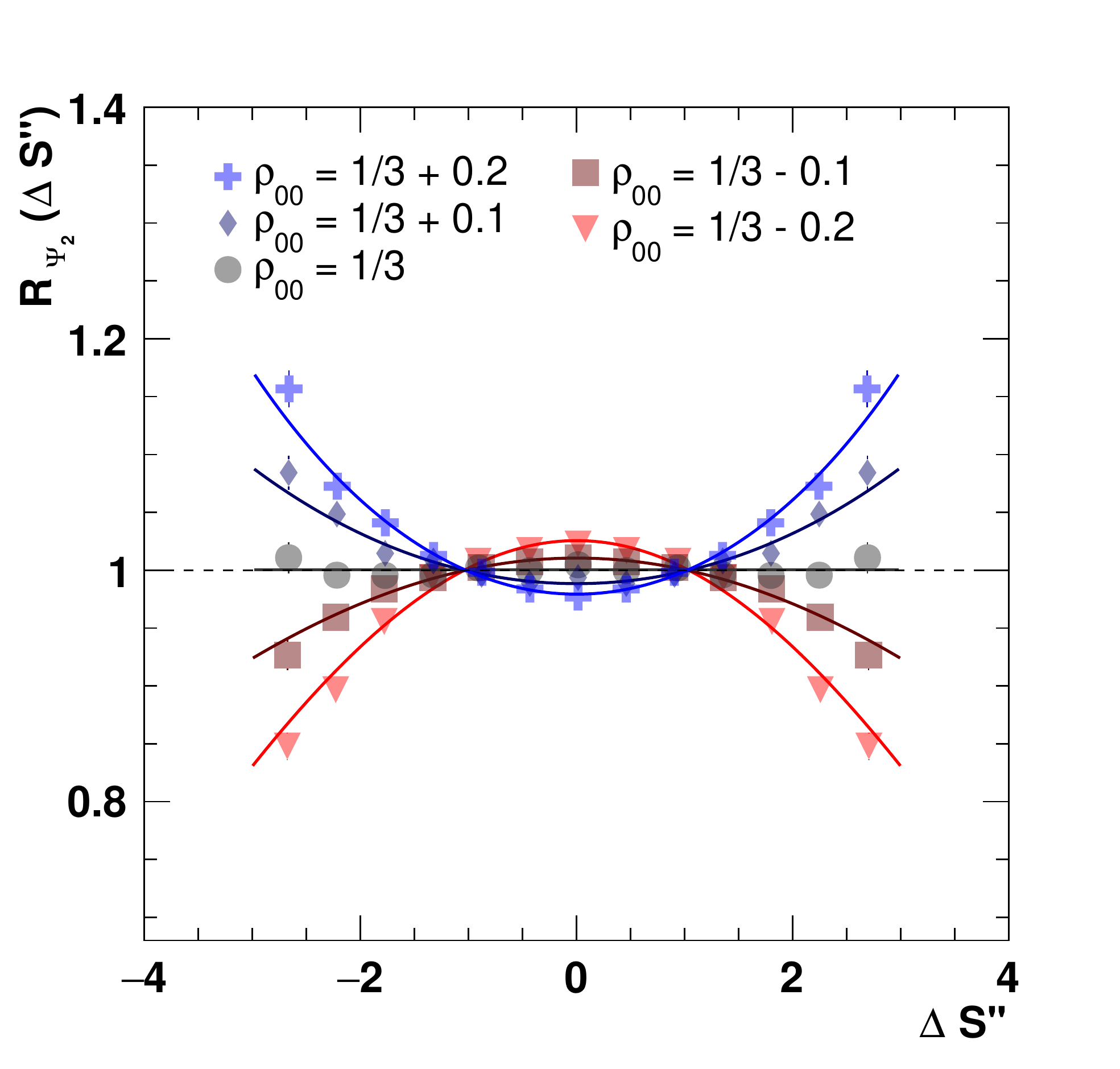}} 	
{\includegraphics[width=0.40\linewidth]{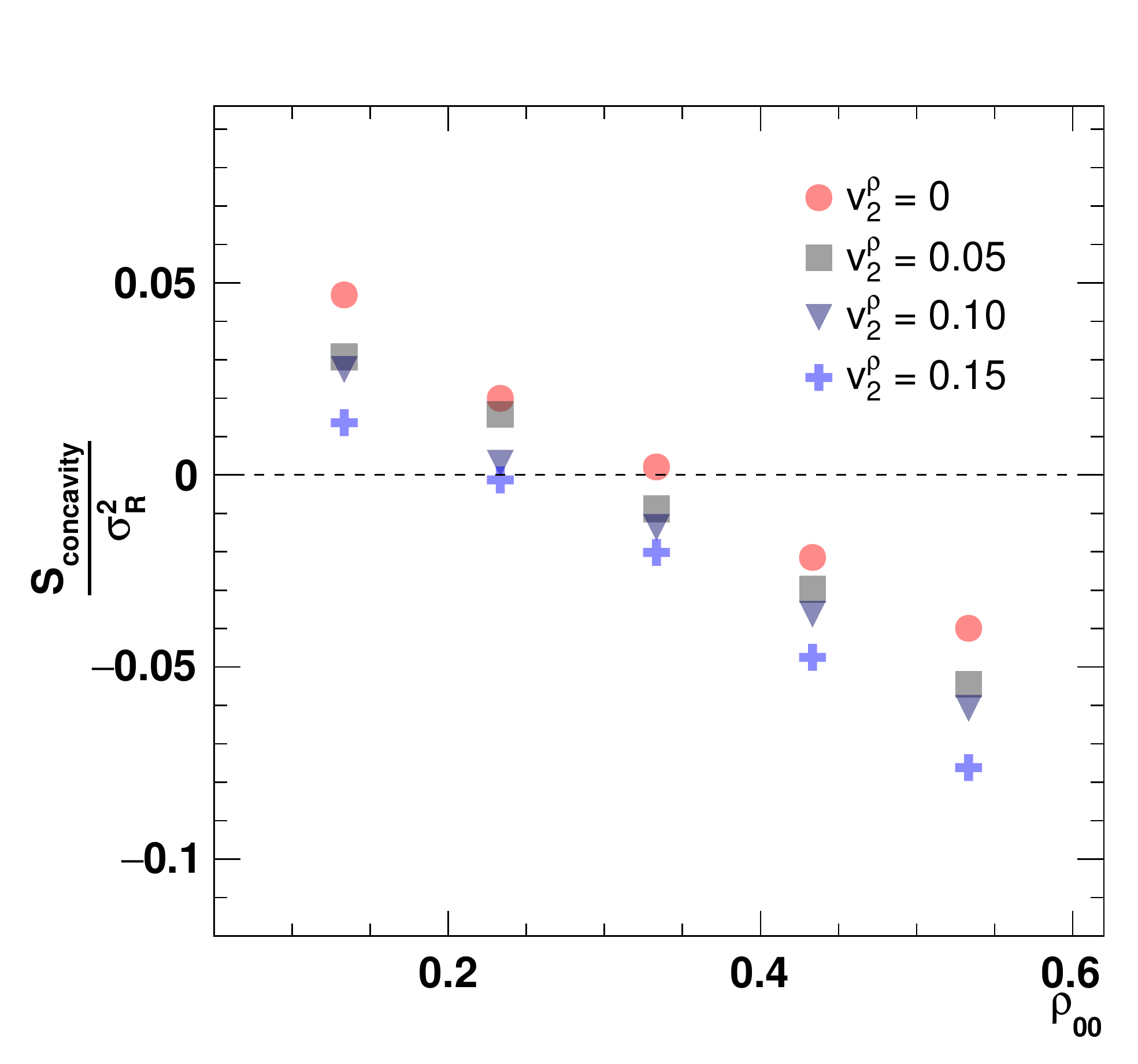}} 
\caption{(Left) toy model simulations of the $R_{\Psi_2}(\Delta S'')$ distribution for events with zero $v_2^\rho$ and different $\rho_{00}$ inputs. The distributions are symmetrized around $\Delta S''=$0.  (Right) $S_{\rm concavity}/\sigma_{R}^2$ extracted using Gaussian fits to $R_{\Psi_2}(\Delta S'')$ for different $v_2^\rho$ and $\rho_{00}$ inputs. A positive  value indicates the convex shape, and a negative one represents the concave shape.}
	\label{fig:RCor_toy}
\end{figure*}
In the following simulations, the generated events and the analysis cuts are the same as in the previous section. 
We take the same procedure as in Ref.~\cite{Magdy:2017yje} to correct the $R_{\Psi_2}(\Delta S)$ correlator for the particle number fluctuations, i.e., $\Delta S'' = \Delta S/\sigma_{\rm sh}$, where $\sigma_{\rm sh}$ is the width of $N(\Delta S_{\rm shuffuled})$.  Figure~\ref{fig:RCor_toy} shows $R_{\Psi_2}(\Delta S'')$ as a function of $\rho_{00}$ from the toy model simulations with zero $v_2^\rho$. The $R_{\Psi_2}(\Delta S'')$ shapes are concave (convex) for $\rho_{00}>1/3$ ($\rho_{00}<1/3)$, indicating a finite background from spin-aligned vector mesons. Figure~\ref{fig:RCor_toy} also shows the $S_{\rm concavity}/\sigma_{R}^2$ extracted using Gaussian fits for different $v_2^\rho$ and $\rho_{00}$ inputs. The red circles represent the case with zero $v_2^\rho$, and corroborate Eq.~(\ref{Eq:26}).  At a given $v_2^\rho$, $S_{\rm concavity}/\sigma_{R}^2$ displays a decreasing trends as a function of $\rho_{00}$. On the other hand, at a given $\rho_{00}$, $S_{\rm concavity}/\sigma_{R}^2$ also decreases with increased $v_2^\rho$, which is consistent with findings in Ref.~\cite{Feng:2018chm}. 

\begin{figure*}[htbp]
	\centering
	\vspace*{-0.1in}
{\includegraphics[width=0.39\linewidth]{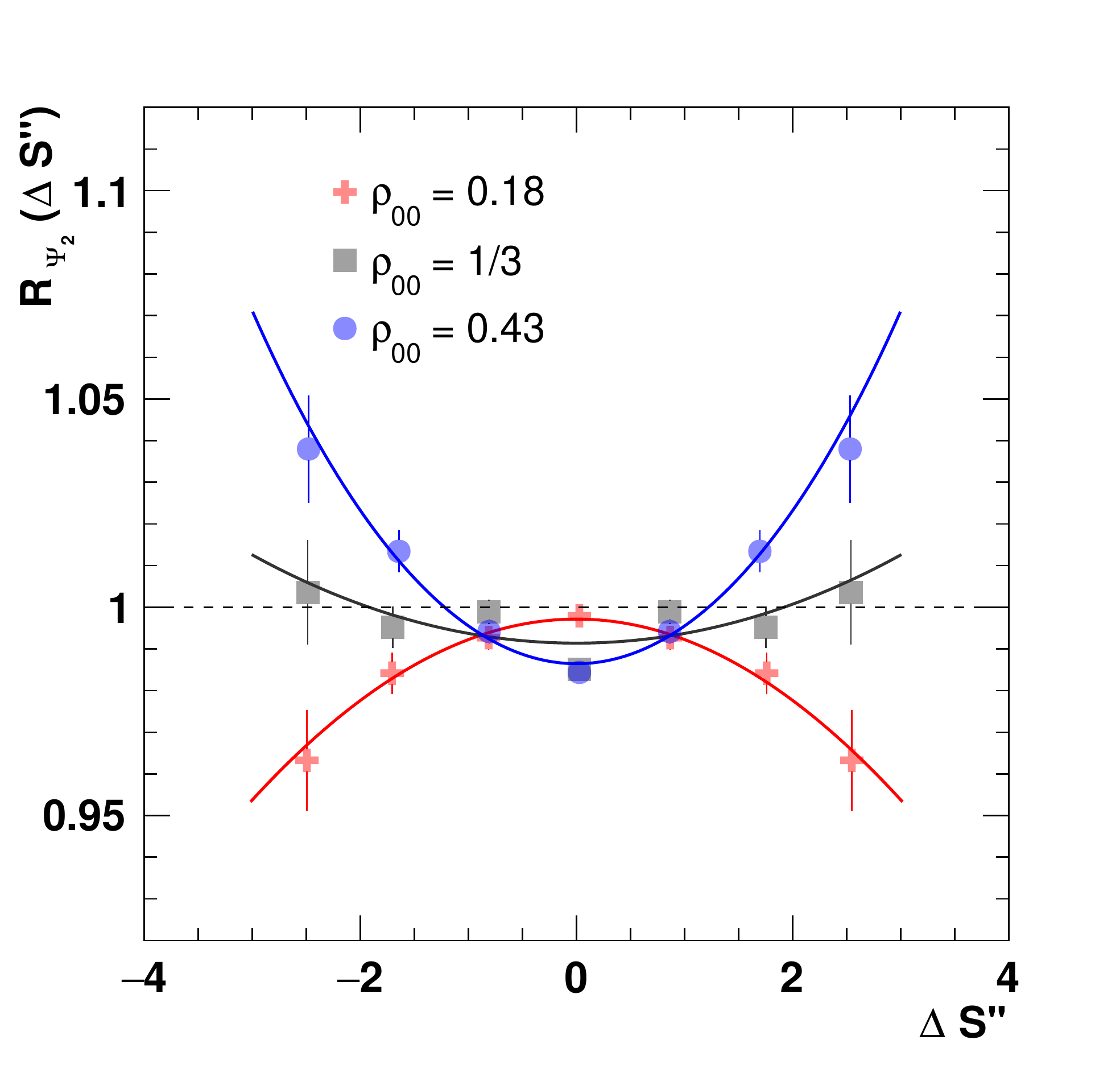}} 
{\includegraphics[width=0.39\linewidth]{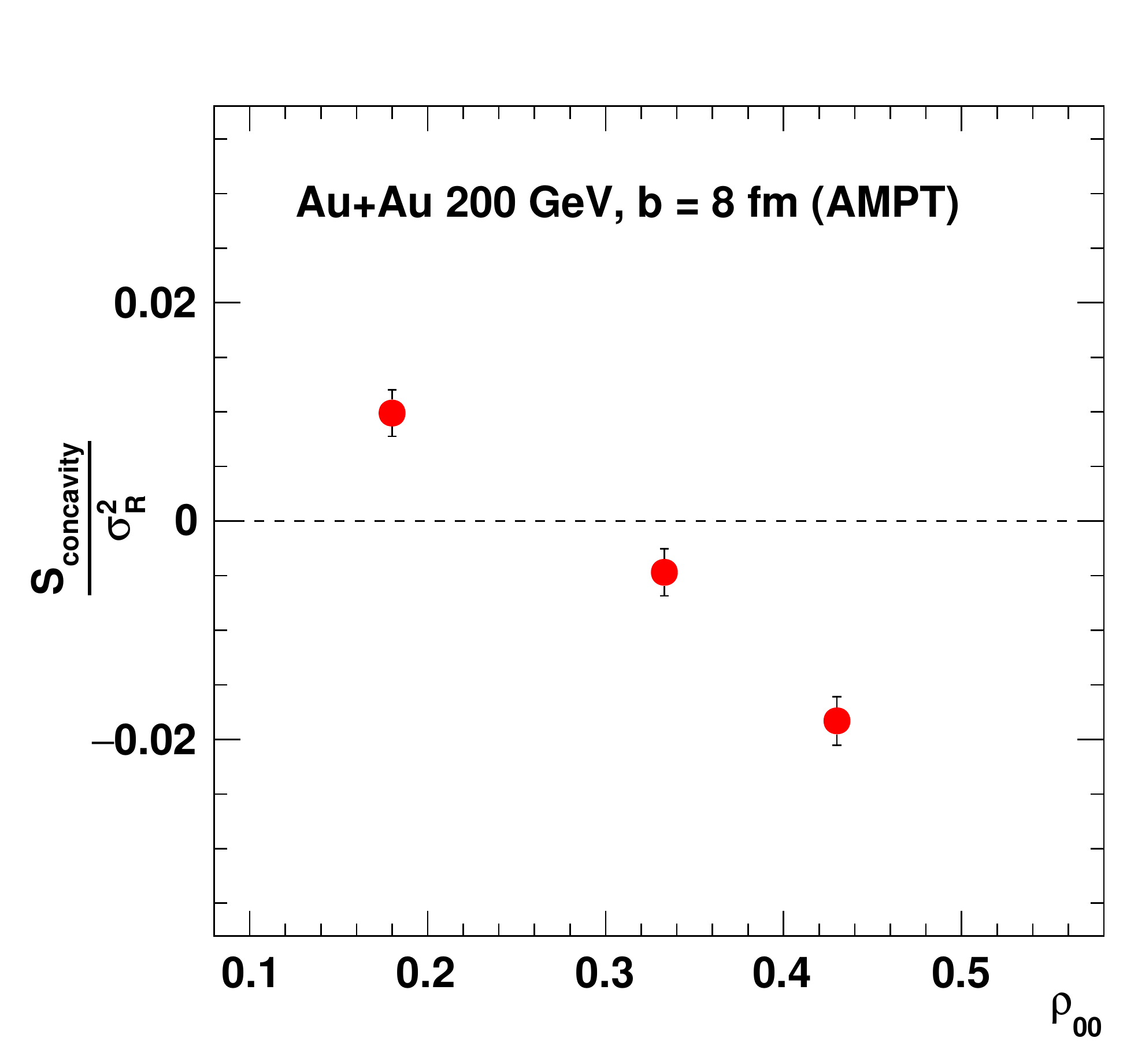}} 
\caption{(Left) AMPT calculations of $R_{\Psi_2}(\Delta S'')$  with $\rho_{00}=$ 0.18 (red cross), 1/3 (gray square) and 0.43 (blue circle) in Au+Au collisions at 200 GeV with the impact parameter of 8 fm. The distributions are symmetrized around $\Delta S''=$0. (Right) $S_{\rm concavity}/\sigma_{R}^2$ extracted using Gaussian fits. }
	\label{fig:RCor_AMPT}
\end{figure*}

Figure~\ref{fig:RCor_AMPT} shows AMPT calculations of $R_{\Psi_2}(\Delta S'')$ with $\rho_{00}=$ 0.18, 1/3 and 0.43, respectively, in Au+Au collisions at 200 GeV with the impact parameter of 8 fm. At $\rho_{00}=1/3$, the concave shape, which suggests a non-CME background, may come from the positive $v_2^\rho$ and transverse momentum conservation. Figure~\ref{fig:RCor_AMPT} also presents the $S_{\rm concavity}/\sigma_{R}^2$ values retrieved using Gaussian fits,
showing a $\rho_{00}$ dependence similar to those from the toy model. Thus the AMPT calculations further confirm the background contribution from globally spin-aligned $\rho$ mesons to the $R_{\Psi_2}(\Delta S)$ correlator.

\begin{figure}[htbp]
	\centering
	\vspace*{-0.1in}
{\includegraphics[scale=0.35]{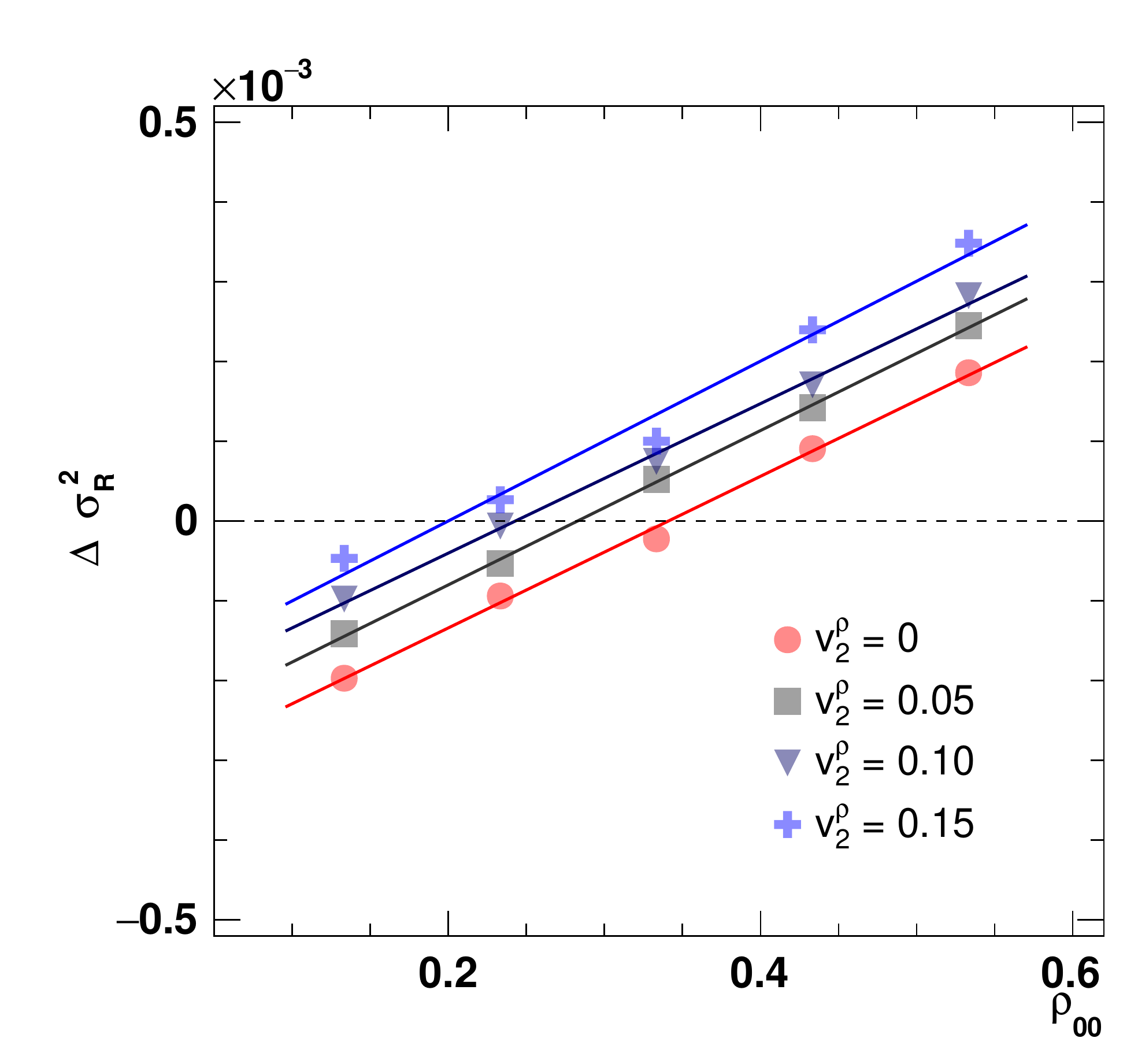}} 	
	\caption{Toy model simulations of $\Delta \sigma_{R}^2$ as a function of $\rho_{00}$ with various  $v_2^\rho$ inputs. Linear fits are applied to guide eyes.}
	\label{fig:RCor_Delta}
\end{figure}

Compared with Eq.~(\ref{Eq:sigma_R2}), it is more appealing to construct the difference variable, $\Delta \sigma_{R}^2$, as proposed in Ref.~\cite{Choudhury:2021jwd}:
\begin{flalign}
&\Delta \sigma_{R}^2 
= \sigma^2(\Delta S_{\rm real}) - \sigma^2(\Delta S_{\rm shuffled}) - \sigma^2(\Delta S_{\rm real}^{\perp}) \notag \\
& + \sigma^2(\Delta S_{\rm shuffled}^{\perp}). &
\end{flalign}
Note that the CME should give a positive contribution to 
$\Delta \sigma_{R}^2$.
The contribution of $\rho$-decay pions to $\Delta \sigma_{R}^2$ is
\begin{flalign}
&\Delta \sigma_{R}^2 = \frac{N_\rho}{N_+N_-} \left [ \frac{1}{4}(f_c+f_s)(3\rho_{00}-1) +(f_c-f_s) \right ].& \label{Eq:RCor_delta}    
\end{flalign}
The function form of Eq.~(\ref{Eq:RCor_delta}) is very similar to Eq.~(\ref{Eq:Gamma_lab}), and $\Delta \sigma_{R}^2$ also has a linear dependence on $\rho_{00}$.
Consequently, the toy model simulations of $\Delta \sigma_{R}^2$ as a function of $\rho_{00}$ with various  $v_2^\rho$ inputs, as depicted in Fig.\ref{fig:RCor_Delta}, are similar to those of $\Delta\gamma_{112}$ in  Fig.~\ref{fig:Gamma_toy}.

\section{The signed balance functions}
The signed balance functions probe the CME by examining the momentum ordering between positively and negatively charged particles, based on the following quantity~\cite{Tang:2019pbl,Choudhury:2021jwd} 
\begin{flalign}
&\Delta B_y  \equiv \Big[\frac{N_{y(+-)}-N_{y(++)}}{N_+} - \frac{N_{y(-+)}-N_{y(--)}}{N_-}\Big] \notag \\
 &  - \Big[\frac{N_{y(-+)}-N_{y(++)}}{N_+} - \frac{N_{y(+-)}-N_{y(--)}}{N_-}\Big] \notag \\
&= \frac{N_+ + N_-}{N_+N_-}[N_{y(+-)} - N_{y(-+)}],&
\label{Eq:Delta_By}	
\end{flalign}
where $N_{y(\alpha\beta)}$ is the number of pairs in which  particle $\alpha$ is ahead of particle $\beta$ along the $y$ axis ($p_y^\alpha > p_y^\beta$) in an event. Similarly, $\Delta B_x$ can be constructed along the $x$ axis. The CME  will enhance the width of the $\Delta B_y$ distribution via the charge separation along the $y$ axis, and therefore the final observable is the ratio
\begin{flalign}
&r \equiv \sigma(\Delta B_y)/\sigma(\Delta B_x).&
\end{flalign}
$r$ can be calculated in both the laboratory frame ($r_{\rm lab}$) and the $\rho$ rest frame ($r_{\rm rest}$). The CME will lead to $r_{\rm rest}>r_{\rm lab}>1$. In this work, we focus on $r_{\rm lab}$, since the extra sensitivity in $r_{\rm rest}$ is a higher-order effect, and requires substantially more statistics and computing resources. 

Ref.~\cite{Tang:2019pbl} has pointed out that the global spin alignment of $\rho$ mesons will affect the $r_{\rm lab}$ measurement. Here, we will derive the qualitative relation. For simplicity, we assume that all the particles have the same $p_T$, and write  $\Delta B_y$ and $\Delta B_x$ as~\cite{Choudhury:2021jwd}  
\begin{flalign}
&\Delta B_y \approx \frac{8M}{\pi^2}\left( 1+\frac{2}{3}\langle \cos2\Delta \phi \rangle \right )\left( \langle \sin\Delta \phi_+ \rangle - \langle \sin\Delta \phi_- \rangle \right), \label{Eq:31} \notag \\
	\\
&\Delta B_x \approx \frac{8M}{\pi^2}\left( 1-\frac{2}{3}\langle \cos2\Delta \phi \rangle \right )\left( \langle \cos\Delta \phi_+ \rangle - \langle \cos\Delta \phi_- \rangle \right), & \label{Eq:32} \notag \\
\end{flalign}
respectively. Here, the bracket means averaging over all particles of interest in an event, and we also assume $N_+\approx N_- =M/2$ and $\langle \cos 2\Delta \phi_+ \rangle = \langle \cos 2\Delta \phi_- \rangle$. After some derivations (see \ref{appendix1} for details), we can relate  $\sigma(\Delta B_y)$ and $\sigma(\Delta B_x)$ to the core components of the $R_{\Psi_2}(\Delta S)$ correlator,
\begin{flalign}
&\sigma^2(\Delta B_y) \approx \frac{64M^2}{\pi^4} \left ( \frac{4}{9M} + 1 + \frac{4}{3}v_2 \right )\sigma^2(\Delta S_{\rm real}), \label{Eq:SigmaBy}\\
&\sigma^2(\Delta B_x) \approx \frac{64M^2}{\pi^4} \left ( \frac{4}{9M} + 1 - \frac{4}{3}v_2 \right )\sigma^2(\Delta S^\perp_{\rm real}).& \label{Eq:SigmaBx} 
\end{flalign}

According to Eqs.~(\ref{Eq:VarYReal}) and (\ref{Eq:VarXReal}), we immediately see the impact of $\rho_{00}$ on $\sigma(\Delta B_y)$ and $\sigma(\Delta B_x)$, as well as their ratio, $r_{\rm lab}$.
However, it is more straightforward to define an observable based on the difference instead of the ratio,
\begin{flalign}
&\Delta \sigma^2(\Delta B) \equiv \sigma^2(\Delta B_y) - \sigma^2(\Delta B_x) \notag \\
&\approx c_1+c_2(3\rho_{00}-1),& \label{Eq:BF_delta}
\end{flalign}
where $c_1$ and $c_2$ are constant coefficients which depend on the spectra of $\rho$ mesons, $v_2^\rho$, and $v_2^\pi$.

\begin{figure}[htbp]
	\centering
	\vspace*{-0.1in}
	\includegraphics[scale=0.35]{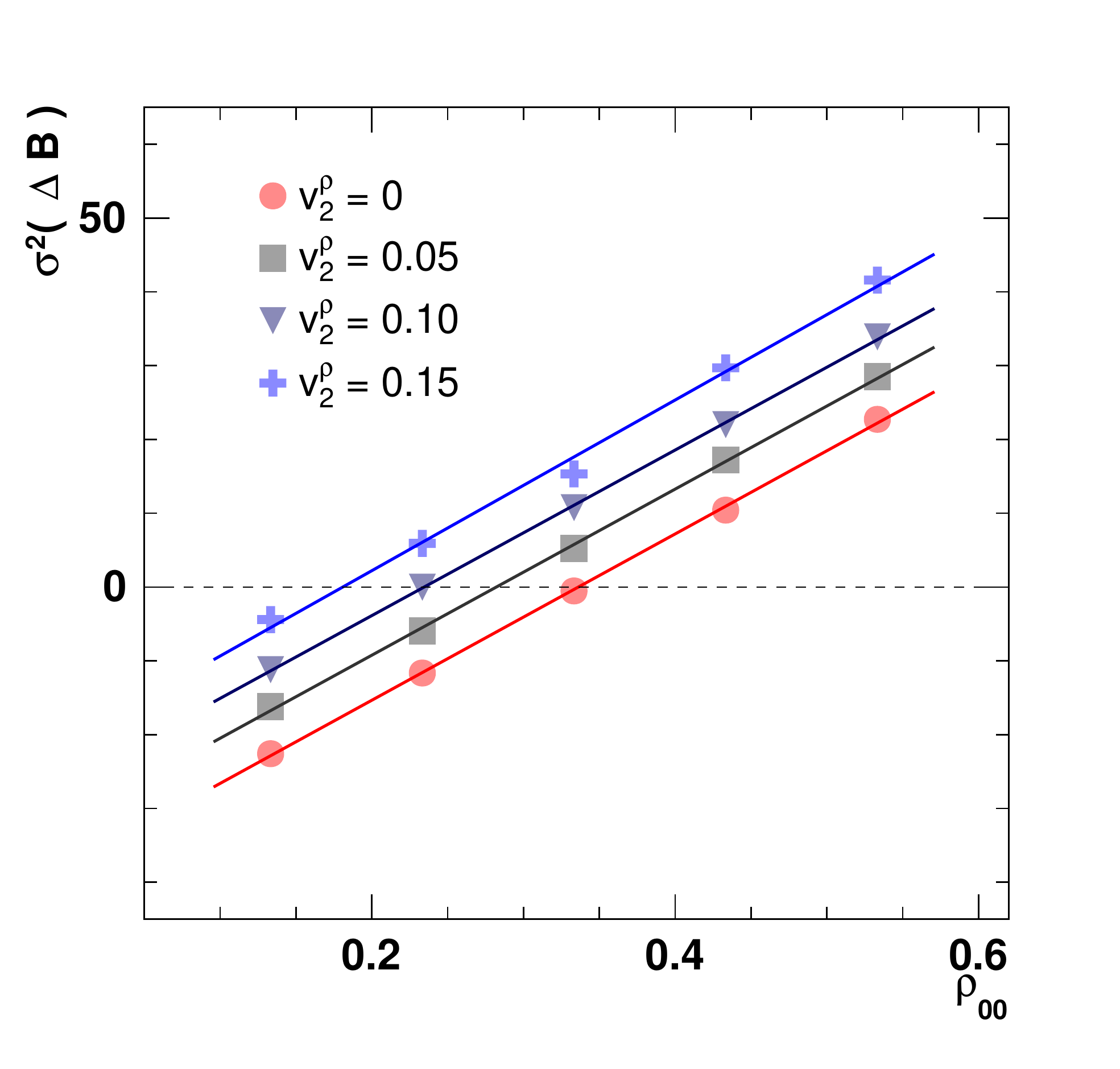}
	\vspace*{-0.2in}
\caption{Toy model simulations of $\sigma^2(\Delta B)$ as a function of $\rho_{00}$ for various  $v_2^\rho$ inputs.  Linear fits are applied to guide eyes.}
	\label{fig:BF_Delta}
\end{figure}

Figure~\ref{fig:BF_Delta} shows the toy model simulations of $\Delta \sigma^2(\Delta B)$ as a function of $\rho_{00}$  with various $v_2^\rho$ inputs. At a given $v_2^\rho$,  $\Delta \sigma^2(\Delta B)$ exhibits a linear dependence on $\rho_{00}$. In the case of $v_2^\rho=0$ and $\rho_{00}=1/3$, $\Delta \sigma^2(\Delta B)$ is zero as expected by Eq.~\ref{Eq:BF_delta}). Note that the global spin alignment effect could give a negative contribution to $\Delta \sigma^2(\Delta B)$, if $\rho_{00}$ is smaller than $1/3$.
The function forms of Eqs.~\ref{Eq:Gamma_lab}), (\ref{Eq:RCor_delta}) and (\ref{Eq:BF_delta}) are very similar to each other, and therefore these observables should have similar sensitivity to the global spin alignment of vector mesons. Indeed, Fig.~\ref{fig:BF_Delta} looks very similar to Figs.~\ref{fig:Gamma_toy} and \ref{fig:RCor_Delta}.

\begin{figure}[htbp]
	\centering
	\vspace*{-0.1in}
{\includegraphics[scale=0.35]{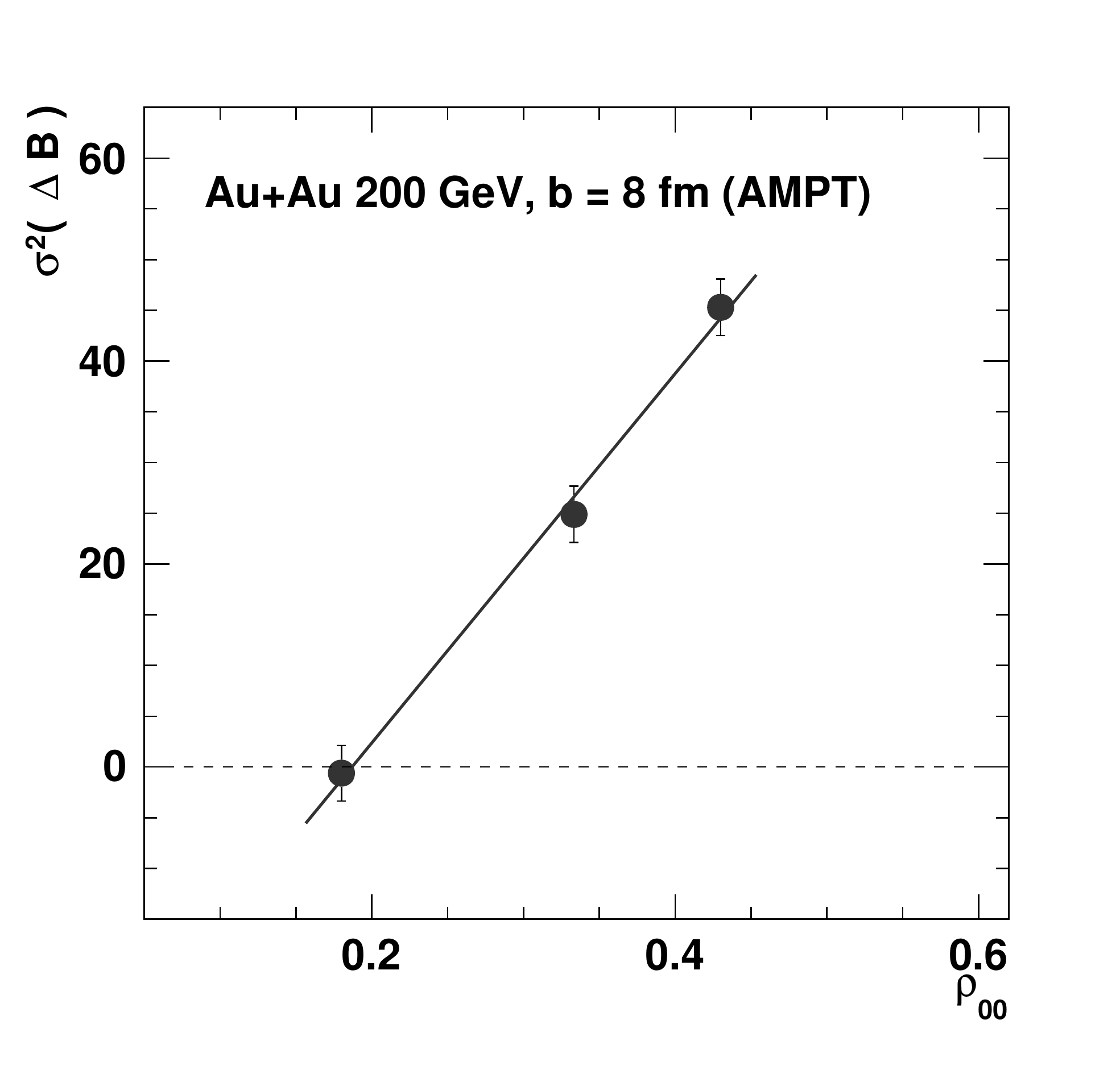}} 	
\caption{AMPT calculations of $\sigma_\rho^2(\Delta B)$ as a function of $\rho_{00}$ in Au+Au collisions at 200 GeV with the impact parameter of 8 fm. A linear fit is applied to guide eyes.}
	\label{fig:BF_DeltaAMPT}
\end{figure}

Figure~\ref{fig:BF_DeltaAMPT} shows AMPT calculations of $\Delta \sigma^2(\Delta B)$ as a function of $\rho_{00}$ in Au+Au collisions at 200 GeV with the impact parameter of 8 fm. $\Delta \sigma^2(\Delta B)$ increases linearly with $\rho_{00}$, similar to the toy model simulations. At $\rho_{00}=1/3$, the non-zero $\Delta \sigma^2(\Delta B)$, which is a non-CME background, may come from the positive $v_2^\rho$ and transverse momentum conservation. 
The slope, $d \Delta \sigma^2(\Delta B)/d\rho_{00}$, could be different between the toy model and the AMPT model, because of the different $\rho$-meson spectra.

\section{Summary}
The chiral magnetic effect in heavy-ion collisions has aroused extensive interest, but the experimental search for the CME is hindered by the incomplete understanding of the background effects. The global spin alignment of vector mesons can provide a sizeable contribution to the CME-sensitive observables that involve the decay products of these vector mesons.
Since the spin alignment could arise from non-CME mechanisms, such as the spin-orbit coupling, it presents a potential background to the CME measurements, in addition to the commonly admitted flow background. 

In this work, we have demonstrated how the globally spin-aligned $\rho$ mesons  affect the CME observables involving pions, the $\Delta \gamma_{112}$ correlator, the $R_{\Psi_2}(\Delta S)$ correlator, and the signed balance functions. For each observable, we first analytically derive its qualitative dependence on $\rho_{00}$, and then confirm the equations with a toy model as well as the more realistic AMPT simulations. Qualitative derivations indicate that the $\rho_{00}$ dependence originates from the anisotropic emission of the decay products, which imitates elliptic flow in the $\rho$ rest frame.
We find that all these observables are influenced not only by  elliptic flow $v_2^\rho$, but also by the spin alignment $\rho_{00}$ of $\rho$ mesons. By constructing new observables using the difference  instead of the ratio for $R_{\Psi_2}(\Delta S)$ and  $r_{\rm lab}$, we have $\Delta \gamma_{112}$, $\Delta \sigma_{R}^2$, and $\Delta \sigma^2(\Delta B)$ all manifest a similar linear dependence on  $\rho_{00}$. Unlike the flow background that is always positive, the global spin alignment of vector mesons can give a negative contribution if $\rho_{00}$ is smaller than 1/3, which is likely according to data. This further warrants the inclusion of the spin alignment effect in the background estimation, to avoid an over-subtraction of the background. 

We cannot exclude the possibility that the global spin alignment partially stems from the CME-induced charge separation of $\pi^+$ and $\pi^-$, some of whom later form $\rho$ mesons via coalescence.
In that case, the CME tends to give a positive contribution to the $\rho_{00}$ of $\rho$ mesons. Such a coupling between the CME and the global spin alignment goes beyond the scope of this work, and calls for more theoretical inputs.

\section*{Acknowledgement}
D. Shen and J. Chen are supported in part by the National Key Research and Development Program of China under Contract No. 2022YFA1604900, by the National Natural Science Foundation of China under Contract No. 12147114, No. 12147101 and No. 12025501, by the Strategic Priority Research Program of Chinese Academy of Sciences under Grant No. XDB34030000, and by the Guangdong Major Project of Basic and Applied Basic Research No. 2020B0301030008. A.H. Tang is supported by the US Department of Energy under Grants No. DE-AC02-98CH10886, DE-FG02-89ER40531. G. Wang is supported by the US Department of Energy under Grant No. DE-FG02-88ER40424.

\appendix

\section{Core components of the Signed Balance Functions}
\label{appendix1}

Based on Eq.~(\ref{Eq:31}), we write the variance of $\Delta B_y$ as 
\begin{flalign}
&\sigma^2(\Delta B_y)
\approx \frac{64M^2}{\pi^4}\sigma^2(x_1x_2),&
\label{Eq:SigmaBy_step1}
\end{flalign}
where $x_1=1+(2/3)\langle \cos2\Delta\phi \rangle$ and $x_2 = \langle \sin \Delta \phi_+ \rangle - \langle \sin \Delta \phi_- \rangle$. Because the cosine and sine functions are orthogonal to each other, $x_1$ and $x_2$ can be treated as independent variables. Therefore, Eq.~(\ref{Eq:SigmaBy_step1}) can be further expanded as
\begin{flalign}
&\sigma^2(\Delta B_y) \approx \frac{64M^2}{\pi^4}[\sigma^2(x_1)\sigma^2(x_2) + \sigma^2(x_1)\langle x_2 \rangle^2 \label{Eq:A2}\notag \\
& + \sigma^2(x_2)\langle x_1 \rangle^2].&
\end{flalign}
We then calculate each term in Eq.~(\ref{Eq:A2}),
\begin{flalign}
&\sigma^2(x_1) = \frac{8}{9M}(\frac{1}{2} + \frac{v_4}{2} - v_2^2),\\
&\sigma^2(x_2) = \sigma^2(\Delta S_{\rm real}), \\
&\langle x_1 \rangle^2 = (1+ \frac{2}{3}v_2)^2, \\
&\langle x_2 \rangle^2 = 0. &
\end{flalign}
We ignore higher-order terms like  $v_4$ and $v_2^2$, and reach
\begin{flalign}
&\sigma^2(\Delta B_y) \approx \frac{64M^2}{\pi^4} \left ( \frac{4}{9M} + 1 + \frac{4}{3}v_2 \right )\sigma^2(\Delta S_{\rm real}). &
\end{flalign}
A similar derivation can be applied to $\sigma^2(\Delta B_x)$ to yield  Eq.~(\ref{Eq:SigmaBx}).

\bibliographystyle{elsarticle-num-names}
\bibliography{ref}

\end{document}